# A novel unsupervised covid lung lesion segmentation based on the lung tissue identification

Faeze Gholamian Khah[1], Samaneh Mostafapour[2], Seyedjafar Shojaerazavi[3], Nouraddin Abdi-Goushbolagh[1], Hossein Arabi[4]

[1]Department of Medical Physics, Faculty of Medicine, Shahid Sadoughi University of Medical Sciences, Yazd, Iran

[2]Department of Radiology Technology, Faculty of Paramedical Sciences, Mashhad University of Medical Sciences, Mashhad, Iran

[3]Department of Cardiology, Ghaem Hospital, Mashhad University of Medical Sciences, Mashhad, Iran

[4]Division of Nuclear Medicine and Molecular Imaging, Department of Medical Imaging, Geneva University Hospital, CH-1211 Geneva 4, Switzerland

**Short title:** Unsupervised covid lesion segmentation

**Abstract**

**Objective:**

Accurate segmentation of lung and infections plays a significant role in quantitative image analysis and the fight against Covid-19. This study aimed to evaluate the performance of a novel unsupervised deep learning-based framework for automated infections lesion segmentation from CT images of Covid-19 patients.

**Methods:**

The datasets used in this study included 500 normal cases and 500 confirmed Covid-19 patients. A residual deep learning network was trained for each dataset separately based on CT images and corresponding lung masks. The novel unsupervised approach proposed for automated infection detection relies on distinguishing healthy lung tissue in normal patients from Covid-induced infection in Covid-19 patients. In the first step, two residual networks were independently trained to identify the lung tissue for normal and Covid-19 patients in a supervised manner (given the manually defined lung masks). These two models, referred to as DL-Covid and DL-Norm for Covid-19 and normal patients, respectively, generate the voxel-wise probability maps for lung tissue identification. To detect Covid lesions, the CT image of the Covid patient is processed by the DL-Covid and DL-Norm models to obtain two lung probability maps. Since the DL-Norm model is not familiar with Covid infections within the lung (owing to the training with only normal subjects), this model would assign lower probabilities to the lesions than the DL-Covid. Hence, the probability maps of the Covid infections could be generated through the subtraction of the two lung probability maps obtained from the DL-Covid and DL-Norm models. Manual lesion segmentation of 50 Covid-19 CT images was used to assess the accuracy of the unsupervised lesion segmentation approach. Different quantitative parameters such as Dice coefficients, Relative Mean HU Diff (%), false-positive and false-negative ratios were computed to evaluate the accuracy of the proposed segmentation methods.

**Results:**

The Dice coefficients of $0.985 \pm 0.003$ and $0.978 \pm 0.010$ were achieved for the lung segmentation of normal and Covid-19 patients in the external validation dataset, respectively. The Relative Mean HU Diff (%) values showed underestimation of the Hounsfield Unit in normal (-2.59 ± 2.371) and Covid-19 (-4.010 ± 2.997) patients. Quantitative results of infection segmentation by the proposed unsupervised method showed the Dice coefficient and Jaccard index of $0.67 \pm 0.033$ and $0.60 \pm 0.06$, respectively. Moreover, a false-positive ratio of $0.072 \pm 0.049$ and a false-negative ratio of $0.062 \pm 0.042$ were observed for the unsupervised lesion segmentation.

**Conclusion:**

Quantitative evaluation of the proposed unsupervised approach for Covid-19 infectious lesion segmentation showed relatively satisfactory results. Since this framework does not require any annotated dataset, it could be used to generate very large training samples for the supervised machine learning algorithms dedicated to noisy and/or weakly annotated datasets.

# 1. INTRODUCTION

Since December 2019, the world has been coping with the global pandemic of the novel Coronavirus Disease (Covid -19) caused by the severe acute respiratory syndrome coronavirus 2 (SARS-CoV-2) [1, 2]. Early diagnosis and treatment of patients with Covid-19 would increase the chance of their survival [3]. Although the reverse-transcription polymerase chain reaction (RT-PCR) test is considered as the standard method for confirming Covid -19 cases, it is reported to have low sensitivity and a high false-negative rate within the first days of infection [4]. However, computed tomography (CT) imaging provides detailed/high-resolution three-dimensional anatomical structures [5-11], exhibiting its effectiveness in detecting radiological patterns associated with Covid-19 including bilateral and peripheral ground-glass opacities, and patchy consolidations in the lung [12]. Therefore, chest CT imaging is considered as a complementary tool with improved sensitivity in the early screening of Covid -19 in clinical practice [13, 14]. Moreover, segmentation of lung and infectious lesions in CT images enables quantitative analysis of lesions [15] and plays a significant role in obtaining further insight into follow-up assessment as well as evaluation of therapy response [1, 16].

Manual delineation of volumetric CT images is a time-consuming task and subject to inter- and intra-observer/expert variability [17]. Adopting an automated lung and lesion segmentation method for Covid-19 patients is challenging due to the low contrast between the lung tissue and the chest wall caused by Covid-19 infections. In addition, high variations in size, texture, and position of pneumonic lesions limit the performance of conventional methods in automated infection segmentation [18]. Recent advances in deep learning models and their powerful feature extraction ability have led to tremendous success in image segmentation [12, 19].

In the current Covid-19 pandemic, several studies have reported the efficient performance of convolutional neural networks such as U-Net, 3D U-Net, U-Net++, and V-Net in lung and lesion segmentation [20]. Nonetheless, the existing approaches are mostly based on the supervised learning scheme [12, 19]. Therefore, a large amount of annotated dataset is required for training the models, which would be expensive, time-consuming, and highly labor-intensive [21]. In addition, high variations in size, texture, and the position of pneumonic lesions add to the complexity of the problem and show the importance of using well-annotated datasets [17]. To address this challenge, some studies concentrated on semi-supervised and/or weakly-supervised approaches to avoid the expensive generation/collection of annotated training datasets [1, 17, 22]. Fan et al. [1] presented a semi-supervised approach based on a randomly selected propagation strategy in which only a few labeled Covid-19 infection images are required to improve the performance of their proposed Inf-Net model. In another study, Yao et al. [22] developed a label-free framework for Covid-19 lesion segmentation in which lesions were synthesized and inserted into normal CT images to generate training pairs.

This study set out to introduce a novel unsupervised lung lesion segmentation from Covid-19 CT images without requiring an annotated dataset. To this end, first, deep learning models were developed for the automated lung segmentation from CT images of normal subjects and patients with Covid-19, namely, DL-Norm and DL-Covid models, respectively. Then, an unsupervised approach was proposed for Covid-19 infectious lesion segmentation to effortlessly generate large annotated datasets for the deep/machine learning models dedicated to the noisy or weakly annotated datasets.

## 2. MATERIALS AND METHODS

### 2.1 Overview of the proposed approach (DL-CovidSeg)

In this approach, two separate/independent models are trained for the lung tissue identification from normal and Covid-19 datasets (DL-Norm and DL-Covid, respectively). The training of the models is performed in a supervised manner to obtain voxel-wise lung probability maps. For the segmentation of Covid lesions, the target CT image undergoes the lung tissue identification through the DL-Norm and DL-Covid models to generate two separate lung probability maps. Since the DL-Norm model is unfamiliar with Covid-induced lesions in the lung, this model would assign low probabilities to these tissues. In contrast, the DL-Covid model (owing to the training with the Covid dataset) would recognize Covid lesions within the lung and assign higher probabilities to them. The differences between the two lung tissue probability maps would be used to estimate the final lesion probability map in the target patient.

Figure 1 shows an overview of the proposed framework. The lung probability maps resulting from the training of the two separate models (DL-Norm, DL-Covid) are shown in Figure 1.A. In Figure 1.B, the lung probability map is generated for the input Covid patient using DL-Norm and DL-Covid models. Finally, the Covid infection probability map is obtained from the subtraction of the two voxel-wise lung probability maps.

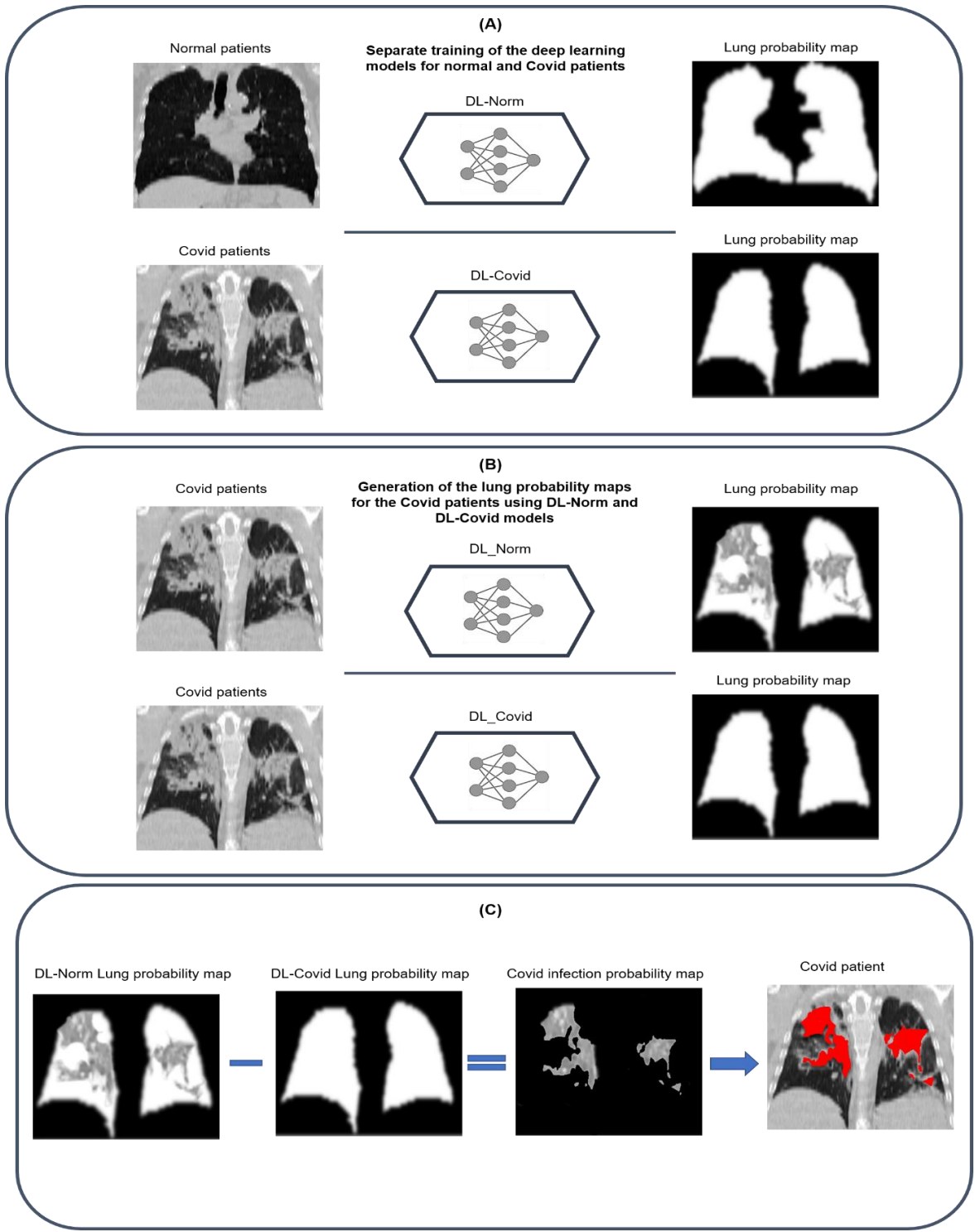

**Figure 1.** Overview of the proposed unsupervised lesion segmentation. (A) Training of the two models for the generation of the lung probability maps for normal and Covid patients. (B) Generation of the lung probability map for the input Covid patient using the two trained models (DL-Covid, and DL-Norm), and (C) the Covid infection probability map obtained from the subtraction of the two lung probability maps.

**2.2 Dataset**

The dataset included in this study consists of 500 normal subjects without any lung involvement and 500 Covid-19 patients with positive RT-PCR. CT images were acquired on a Siemens Somatom Spirit Dual Slice CT scanner with the following parameters: tube energy of 130 kVp, tube current of 48 mAs, rotation time of 0.8 s, and slice thickness of 5 mm. For generating ground truth lung masks, semi-automated segmentation of CT images was performed via Toolkit (PTK) software [23] and the resulting binary masks were checked and refined manually if necessary.

Given the lung masks, all images were cropped using the lung bounding box to eliminate the irrelevant background, and they were resized to 296×216 matrix size using a linear interpolation algorithm. To reduce the dynamic range of image intensities, the CT values (Hounsfield Unit (HU)) between -1000 HU and 200 HU were normalized between 0 and 1. An experienced radiologist manually segmented Covid lesions (infections) within the lung for 50 subjects in the external validation dataset. These manual delineations were utilized to assess the performance of the unsupervised lesion detection.

**2.2 Implementation details**

In this study, the automated lung segmentation was performed using a ResNet model implemented in the NiftyNet [24]. The ResNet architecture, as represented in Figure 2, consists of 20 convolutional layers in which every two layers are linked together by a residual connection, followed by a fully connected SoftMax layer as the last layer of the network. For extracting low-level, mid-level, and high-level features from the input image, the convolutional kernels of the network are dilated by the factors of one, two, and four, respectively.

**Lung mask prediction:** The supervised training of the deep learning models was performed separately for normal and Covid-19 datasets. To this end, 450 pairs of CT images and reference lung masks of each dataset were selected for training, and 50 CT images were devoted to the external validation. Also, to avoid any potential overfitting, 5% of the training datasets were employed for validating the model during the training. The same parameters were used for the training of both models as follows: spatial window size: 296×216×1, learning rate = 0.02 - 0.001, optimizer = Adam, loss function = Dice_NS, decay = 0.0001, batch size = 20, and weights regression type = L2norm.

It should be noted that the output of the DL-Covid (trained only with Covid subjects) and DL-Norm (trained only with normal subjects) models, as shown in Figure 1A, are voxel-wise lung probability maps. The larger the value assigned to a voxel, the more likely the voxel belongs to the intrapulmonary area. Accordingly, a lung mask can be created by setting a threshold value to distinguish between the lung and the background tissues.

**Lesion mask prediction**: The proposed unsupervised approach for the infection prediction relies on the fact that there are no lesions inside healthy lungs. In this regard, as shown in Figure 1.B, DL-Covid and DL-Norm models were separately applied to the CT images of Covid-19 subjects. Since the DL-Covid was trained by the dataset of Covid-19 patients, this model accurately included the Covid-induced lesions in the lung regions. In fact, the DL-Covid model gives high probabilities to the lesions to be considered as the lung tissue [25]. On the contrary, the DL-Norm model, which was trained with only normal subjects and is only familiar with healthy/normal lung tissue, would assign low values/probabilities to the Covid lesions to be considered as the lung tissue. The subtraction of the lung probability maps obtained from the DL-Covid and DL-Norm models for the same subject would highlight the areas with Covid infection (Figure 1.C). The probability map obtained from the subtraction (absolute values were considered to avoid the negative numbers) of the two DL-Covid and DL-Normal probability maps was employed to delineate the infected area. To this end, the receiver operating characteristic was plotted for the 50 subjects in the external validation set versus the manually defined reference lesion masks. An empirical threshold of 0.3 was optimized on the obtained probability map to achieve the best Covid lesion delineation.

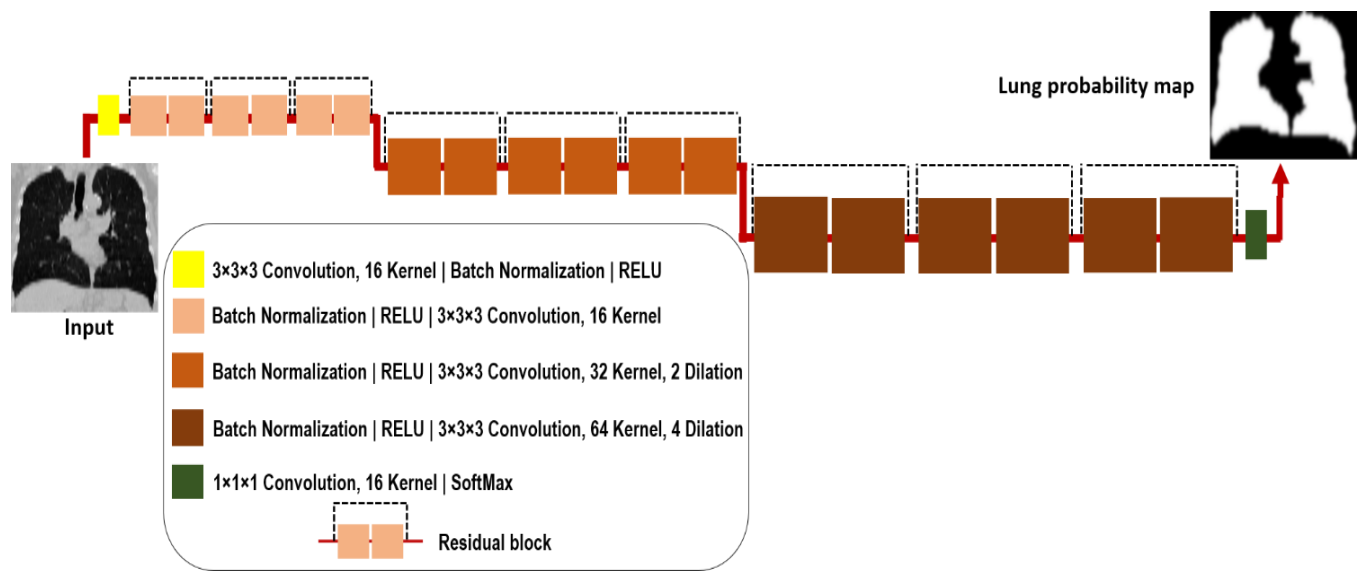

**Figure 2.** The architecture of the deep residual neural network (ResNet).

## 2.3 Evaluation Metrics

The performance of the proposed lung segmentation method was evaluated by comparing the predicted lung segmentation against the reference lung masks in normal and Covd-19 external validation datasets. The assessment was accomplished through calculating dice similarity coefficient (Dice) (Eq. 1) and Jaccard index (JC) (Eq. 2) for the lung as follows:

$$\text{Dice} = \frac{2 \mid L_{ref} \cap L_{pred} \mid}{\mid L_{ref} \mid + \mid L_{pred} \mid} \tag{1}$$

$$\text{JC} = \frac{\mid L_{ref} \cap L_{pred} \mid}{\mid L_{ref} \cup L_{pred} \mid} \tag{2}$$

Here, $L_{ref}$ and $L_{pred}$ indicate the reference and the predicted lung volumes, respectively.

Moreover, relative mean CT number (Hounsfield Unit (HU)) difference, absolute relative mean HU difference, relative volume difference, and absolute relative volume difference parameters were calculated between the predicted and the reference lung masks.

To validate the performance of the lesion segmentation approach, Dice, JC, average Hausdorff distance, mean surface distance (Eq. 3), false-negative ratio (Eq. 4), and false-positive ratio (Eq. 5) metrics were estimated between the predicted and the ground truth lesion masks across 50 Covid-19 patients.

$$\text{Mean Surface Distance} = \frac{1}{2} [\, \bar{d}(S_{pred}, S_{ref}) + \bar{d}(S_{ref}, S_{pred})\,] \tag{3}$$

$$\text{False Negative Ratio} = \frac{FN}{FN + TP} \tag{4}$$

$$\text{False Positive Ratio} = \frac{FP}{FP + TN} \tag{5}$$

In Eq. 3, $\bar{d}(S_{pred}, S_{ref})$ is the mean of distances between each surface voxel in the predicted lesion mask and the closest surface voxel in reference to lesion masks, respectively. In Eq. 4, *FN* and *TP* denote the number of false negatives and true positives, and in Eq. 5, *FP* and *TN* are the numbers of false positives and true negatives, respectively.

A receiver operating characteristic (ROC) curve was generated to display the overall efficiency of the proposed unsupervised approach. Moreover, lesion to lung volume ratio and relative volume differences were calculated between the predicted and the ground truth lesions.

## 3. RESULTS

Figure 3 depicts representative results of lung segmentation for a normal and a Covid patient, as well as the Covid infection identification in the coronal view. The second row shows the predicted lung probability maps from DL-Norm and DL-Covid models for a Covid patient. The Covid infection probability map and the reference infection segmented by the specialist are shown in Figure 3C. There is a good agreement between the ground truth and the unsupervised lesion segmentation.

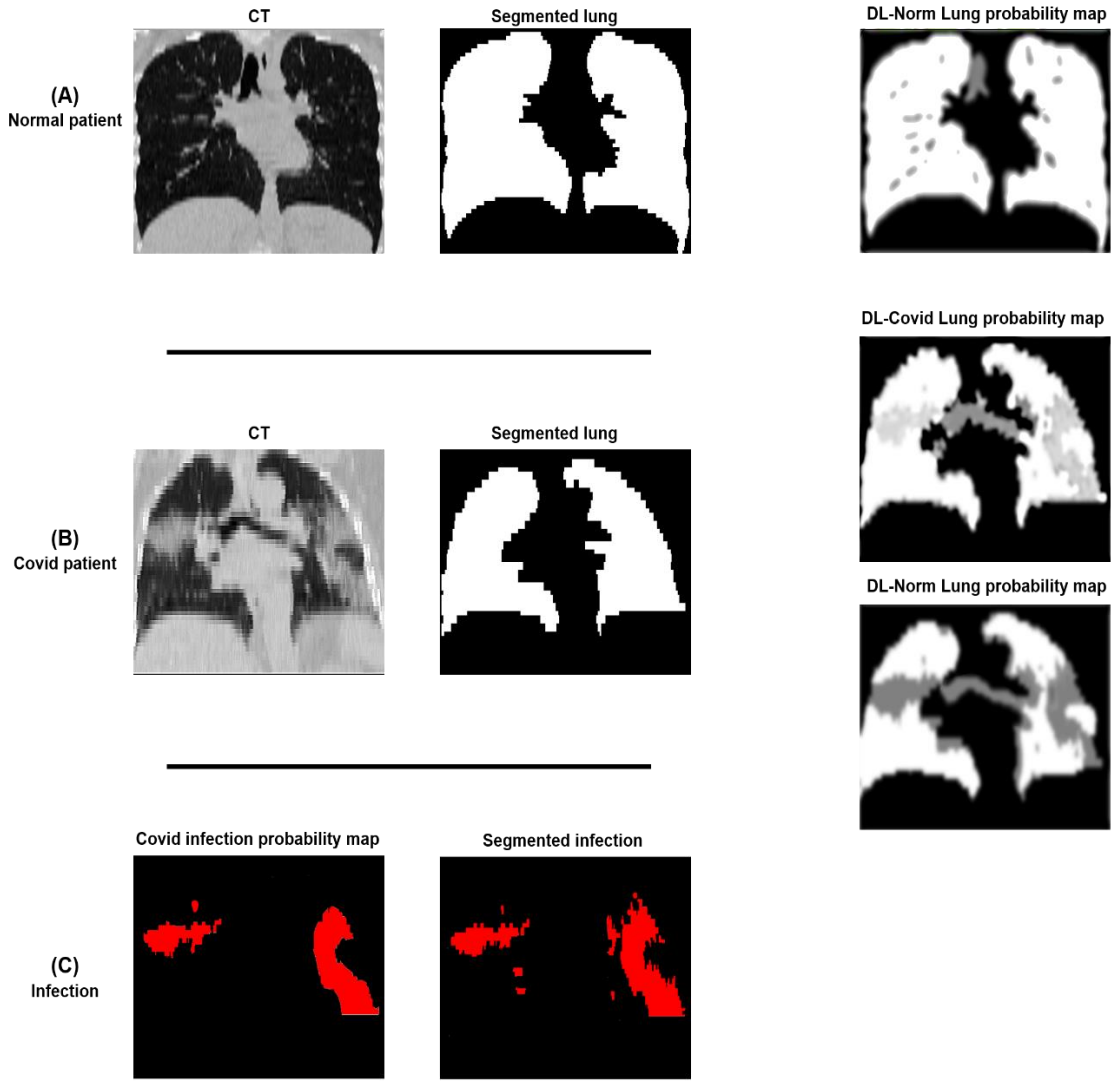

**Figure 1.** Representative coronal views of the CT images, lung segmentation, and the lung probability maps generated by DL-Norm for a normal patient (A) DL-Covid, and DL-Norm for a Covid patient (B). The resulting Covid infection probability map and the reference infection segmented by the specialist are shown in (C).

The results of the lung segmentation evaluation in the external validation set are provided in Table 1. The mean Dice coefficients of 0.985 ± 0.003 and 0.978 ± 0.010 were obtained for the lung segmentation of normal and Covid-19 patients, respectively. The mean Jaccard indices of 0.965 ± 0.006 and 0.943 ± 0.033 were achieved for the lung segmentation of normal and Covid-19 groups, respectively. These results revealed that the proposed method had a slightly better performance in lung segmentation for normal subjects than the Covid-19 patients. The Relative Mean HU Diff (%) values showed underestimation of the Hounsfield Unit in normal (-2.59 ± 2.371%) and Covid-19 (-4.010 ± 2.997%) patients. The proposed method resulted in a Relative Volume Diff (%) of 2.006 ± 6.443 and 4.123 ± 8.652 in the segmentation of the lung tissue for normal and Covid-19 subjects, respectively.

**Table 1.** Quantitative comparison of lung segmentation for normal and Covid-19 subjects.

|  | Dice | JC | Relative Mean HU Diff (%) | Absolute Relative Mean HU Diff (%) | Relative Volume Diff (%) | Absolute Relative Volume Diff (%) |
|---|---|---|---|---|---|---|
| **Normal** | 0.985 ± 0.003 | 0.965 ± 0.006 | -2.59 ± 2.371 | 3.828 ± 1.318 | 2.006 ± 6.443 | 6.766 ± 5.050 |
| **Covid_19** | 0.978 ± 0.010 | 0.943 ± 0.033 | -4.010 ± 2.997 | 4.23 ± 1.897 | 4.123 ± 8.652 | 9.898 ± 7.486 |

Figure 4 presents examples of the Covid-19 infected regions in CT axial slices of four different patients and the segmented infection regions resulting from the proposed unsupervised method (DL-CovidSeg).

Furthermore, the receiver operating characteristic (ROC) curve of the proposed method is plotted in Figure 5, which indicates an area under the curve (AUC) of 0.68 for this unsupervised method.

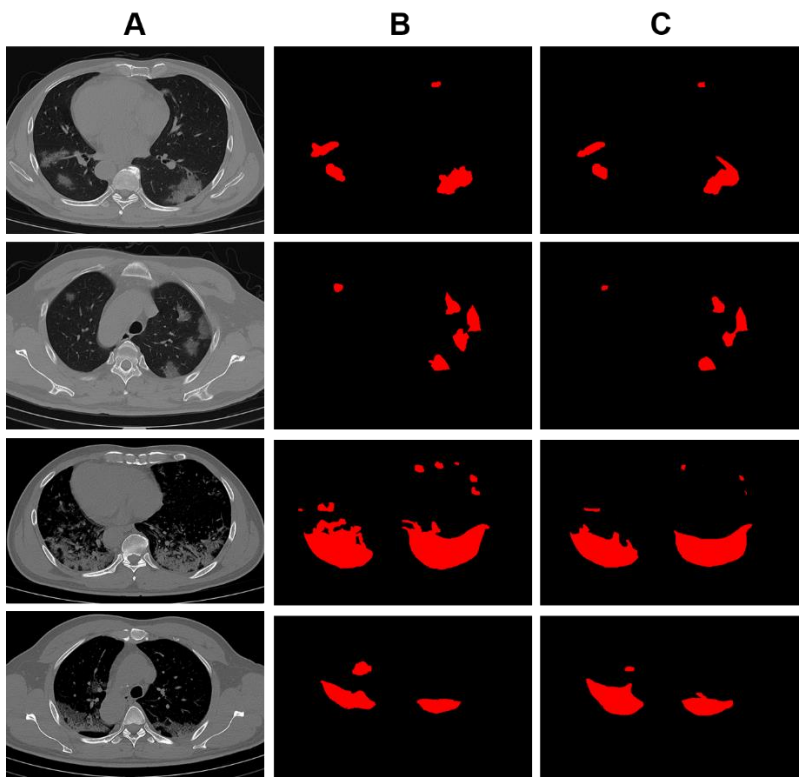

**Figure 2.** Illustrative examples of (A) the Covid-19 infected regions in CT axial slice of four different patients, (B) the ground-truth infection regions segmented by the specialist, and (C) the segmented infection regions resulting from the DL-CovidSeg.

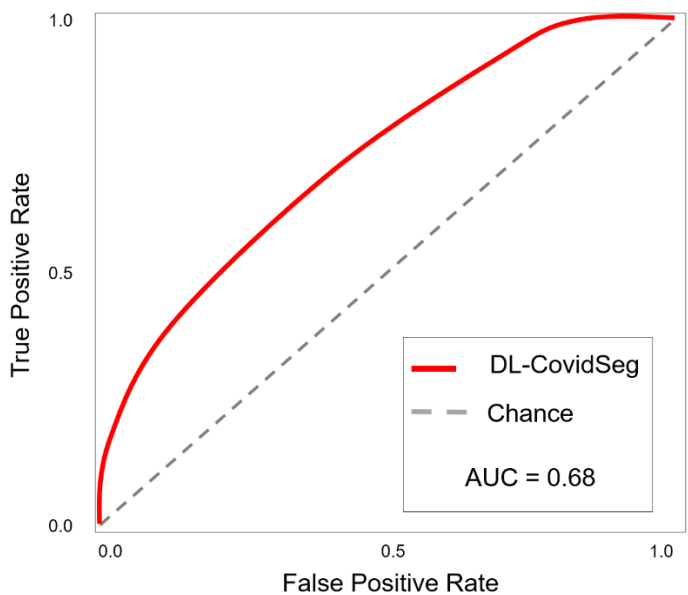

**Figure 3.** The receiver operating characteristic (ROC) curve plotted for the Covid infection probability maps. (AUC: Area under the ROC curve).

Table 2 presents quantitative results of infection segmentation across 50 Covid-19 patients in the external validation set. The Dice coefficient and Jaccard index of 0.67 ± 0.033 and 0.60 ± 0.06 were achieved by the DL-CovidSeg approach, respectively. The false-positive ratio of 0.072 ± 0.049, false-negative ratio of 0.062 ± 0.042, Average Hausdorff Distance of 0.99 ± 0.88, and Mean Surface Distance of 1.1 ± 0.86 were obtained from the unsupervised lesion segmentation approach. Moreover, the relative volume indices are presented in Table 3.

**Table 2.** Statistics of quantitative metrics calculated between reference and predicted infection segmentation.

|  | Dice | JC | False Negative | False Positive | Average Hausdorff Distance | Mean Surface Distance |
|---|---|---|---|---|---|---|
| **Lesion** | 0.67 ± 0.033 | 0.60 ± 0.061 | 0.062 ± 0.042 | 0.072 ± 0.049 | 0.99 ± 0.88 | 1.1 ± 0.86 |

**Table 3.** Relative volume indices for the reference and predicted lesions.

|  | Min | Max | Mean±SD |
|---|---|---|---|
| **Relative lesion/lung volume (manual segmentation)** | 0.001 | 0.75 | 0.14 ± 0.20 |
| **Relative lesion/lung volume (DL-CovidSeg)** | 0.002 | 0.82 | 0.17 ± 0.21 |
| **Relative error of estimated lesion volume (%)** | -11 | 22 | 3.42 ± 7.2 |
| **Absolute relative error of estimated lesion volume (%)** | 0.1 | 22 | 7.9 ± 5.5 |

## 4. DISCUSSION

Image segmentation is an important process for clinical judgment in various diseases since it can make their early diagnosis and effective treatment possible [26-29]. For instance, lung segmentation is crucial in different pathological situations such as trauma, lung cancer, chronic obstructive pulmonary disease (COPD), etc. [30-32]. After the outbreak of the Covid-19 disease in 2019, lung segmentation of CT images has received more importance [33]. Given that, the manual segmentation method is a tedious, time-consuming task that needs clinical experts. In recent years, various segmentation approaches, including thresholding-based methods [34], region-based methods [35], shape-based methods (atlas-based and model-based) [36-38], and neighboring anatomy-guided methods [39] have been proposed in lung CT images; however, they have faced numerous challenges in the segmentation of abnormal lung regions [40]. Recently, deep learning-based methods have shown promising performance in image segmentation, particularly in lung segmentation of patients infected with Covid-19 [41], but most of them require an enormous amount of annotated training dataset of lung lesions provided by specialists, which is a costly and time-consuming task [42]. Thus, some unsupervised or semi-supervised segmentation methods, especially for Covid-19 infection detection, have been proposed to alleviate this issue. For example, Fan et al. [1], in a study, proposed a novel Covid-19 lung infection segmentation deep network (Inf-Net) to automatically recognize the infected regions in CT images. They developed the Inf-Net on top of a semi-supervised learning procedure (Semi-Inf-Net) to increase the number of training subjects; hence, enhancing the performance/robustness of the supervised network. They reported Dice coefficients of 0.682, 0.739 for the proposed Inf-Net and Semi-Inf-Net, respectively. In the present study, we implemented lung segmentation of normal and Covid-19 subjects using a supervised model by the ResNet network. Moreover, we have also developed an unsupervised segmentation procedure for the infected regions in Covid-19 CT images without using any labeled training dataset, wherein comparable results (Dice of 0.67) to the Inf-Net model were observed.

There are several deep learning-based lung segmentation and lesion algorithms, especially using the U-Net network, proposed for Covid-19 patient management. Müller et al. [15] executed a 3D U-Net architecture for lung and infection segmentation through a 5-fold cross-validation scheme on 20 labeled Covid-19 CT images. They utilized extensive data augmentation to reduce the risk of overfitting on their limited data and achieved Dice coefficients of 0.95 and 0.761 for lung and lesion segmentation, respectively. Trivizakis et al. [43] proposed a deep learning model for distinguishing Covid-19 against community-acquired pneumonia. They used a U-Net architecture for lung segmentation that was trained and evaluated on the LIDC-IDRI CT dataset and achieved a Dice Similarity Coefficient of 99.55%. Yu Qiu et al. [44] utilized a specially designed architecture as an extremely minimal network named MiniSeg for the segmentation of Covid-19 infections from chest CT images. This model relied on a small number of trainable parameters and resulted in a Dice Similarity Coefficient of 76.27%. However, all these studies are based on supervised deep learning models.

Here, we performed the segmentation of Covid-19 lung infection from CT images without using any annotated data, with a Dice coefficient of $0.67 \pm 0.033$, which can then be employed for supervised deep learning frameworks dedicated to noisy or weakly annotated datasets. Yao et al. [45] presented a label-free method (NormNet) to distinguish Covid-19 lesion from healthy tissues by generating synthetic 'lesions' and embedding them into normal CT images. They trained a NormNet (with 3D U-Net as the backbone) to assign high confidence to the healthy parts and low confidence to the lesion parts of the volume. They achieved the Dice coefficient of $65.5 \pm 17.9$, and $54.2 \pm 17.5$ for the two groups of data set, Coronacases and Radiopedia, respectively. Although these models exhibited inferior quantitative results compared to the supervised models, the fact that these models do not require labeled datasets shows that they could partly address the issue of large annotated (training) data generation.

## 5. CONCLUSION

In this study, we proposed a novel unsupervised deep learning-based model for the segmentation of Covid-19 infections from CT images. Since this framework does not require any annotated dataset, it could be used to generate large training samples for the supervised machine learning algorithms dedicated to noisy and/or weakly annotated datasets.

**Compliance with ethical standards**

Disclosure of potential conflicts of interest: none of the authors have affiliations that present financial or non-financial competing interests for this work.
Research involving human participants: All procedures performed in studies involving human participants followed the ethical standards of the Mashhad University of Medical Science.
Informed consent: informed consent was obtained from all individual participants included in the study.